\documentclass[11pt,english]{article}
\usepackage{ae,aecompl}
\usepackage[T1]{fontenc}
\usepackage[latin9]{inputenc}
\usepackage{geometry}
\usepackage{babel}
\usepackage{amsmath}
\usepackage{amsthm}
\usepackage{amssymb}
\usepackage{setspace}
\usepackage[unicode=true,pdfusetitle,
 bookmarks=true,bookmarksnumbered=false,bookmarksopen=false,
 breaklinks=false,pdfborder={0 0 1},backref=false,colorlinks=false]
 {hyperref}

\makeatletter
\theoremstyle{plain}
\newtheorem{prop}{\protect\propositionname}
\theoremstyle{plain}
\newtheorem{lem}{\protect\lemmaname}

\usepackage{ae,aecompl}
\usepackage{amssymb}
\usepackage{amsmath}
\usepackage{verbatim}
\usepackage{url}
\usepackage{enumerate}
\usepackage{ifthen}
\usepackage[svgnames]{xcolor}
\usepackage{graphicx}
\usepackage{caption}
\usepackage{subcaption}
\usepackage{booktabs}
\usepackage{array}
\usepackage{multirow}
\usepackage{graphicx,caption}

\usepackage{tikz}

\def\refe #1\par{\noindent\hangindent =\parindent
\hangafter =1 #1\par}

\hypersetup{
  colorlinks   = true, 
  urlcolor     = blue, 
  linkcolor    = blue, 
  citecolor   = blue 
}

\makeatother

\providecommand{\lemmaname}{Lemma}
\providecommand{\propositionname}{Proposition}

\begin{document}
\title{\textbf{``Soft'' Affirmative Action and }\\
\textbf{Minority Recruitment}\thanks{We thank Mariacristina De Nardi, Matthew Notowidigdo, Wojciech Olszewski,
Nicola Persico, Analia Schlosser, and Marciano Siniscalchi for useful
comments and suggestions.}}
\date{April 2020}
\author{Daniel Fershtman\thanks{Eitan Berglas School of Economics, Tel-Aviv University. Email: danielfer@tauex.tau.ac.il}
\and Alessandro Pavan\thanks{Department of Economics, Northwestern University. Email: alepavan@northwestern.edu}}
\maketitle
\begin{abstract}
\begin{onehalfspace}
We study search, evaluation, and selection of candidates of unknown
quality for a position. We examine the effects of \textquotedblleft soft\textquotedblright{}
affirmative action policies increasing the relative percentage of
minority candidates in the candidate pool. We show that, while meant
to encourage minority hiring, such policies may backfire if the evaluation
of minority candidates is noisier than that of non-minorities. This
may occur even if minorities are at least as qualified and as valuable
as non-minorities. The results provide a possible explanation for
why certain soft affirmative action policies have proved counterproductive,
even in the absence of (implicit) biases.
\end{onehalfspace}

\bigskip{}
\bigskip{}
\end{abstract}

\newpage{}

\section{Introduction}

In 2003, the National Football League established the ``Rooney Rule'',
a policy requiring teams to interview minority candidates for head
coaching vacancies.\footnote{See \hyperlink{DuBois}{DuBois (2015)} for an empirical assessment
of the Rooney Rule's impact.} This policy, versions of which have been applied across various industries\footnote{Facebook recently adopted a similar policy (https://money.cnn.com/2018/05/31/technology/facebook-board-diversity/index.html).
In 2014, Senate Resolution (511) was introduced (though ultimately
not enacted) to encourage companies to voluntarily establish policies
to identify and interview qualified minority candidates for managerial
openings at the director level or above.}, is an example of ``soft'' affirmative action (SAA), a term typically
referring to policies designed to change the composition of the candidate
pool, rather than the criteria used during the hiring process. Contrary
to ``hard'' affirmative action policies requiring direct consideration
of minority status as a part of the hiring decision (e.g., employment
quotas)\footnote{Title VII of the Civil Rights Act of 1964 prohibits firms from establishing
\textquotedblleft hard\textquotedblright{} affirmative-action policies
that require direct consideration of minority status during the hiring
process.}, such policies often involve taking steps to increase the share of
minority candidates considered for a position, but subsequently treating
candidates impartially; they are also far less controversial (\hyperlink{Schuck}{Schuck, 2002}).

SAA policies are also common in academic faculty recruitment. For
example, Columbia University's ``Guide to Best Practices in Faculty
Search and Hiring'' offers a checklist of best practices in search
and evaluation in tenure and tenure-track recruitment, which includes:
creating a search plan including broader outreach; signaling a special
interest in candidates contributing to the department\textquoteright s
diversity priorities; reaching out to colleagues to identify high-potential
female and underrepresented minority candidates and encouraging them
to apply to the position; requiring diversity statements; requiring
an explanation (to the Dean) for being unable to find a sufficient
number of competitive candidates from underrepresented groups, and
detailing what steps were taken to identify such candidates.\footnote{See https://provost.columbia.edu/sites/default/files/content/BestPracticesFacultySearchHiring.pdf.}

This paper studies the effectiveness of SAA policies on minority recruitment.
Indeed, while some SAA policies have proven successful (\hyperlink{Heilman}{Heilman, 1980}),
others, such as the Rooney Rule, have been deemed ineffective, or
even counterproductive. One possible explanation is the presence of
implicit biases. Our analysis provides an alternative explanation
based on the well-established observation that the evaluation of minority
candidates is often noisier than that of non-minorities. We show that
this property alone may be responsible for the negative effects of
SAA on minority recruitment. We show that, while meant to increase
the likelihood of hiring a minority candidate, such policies may in
fact lead to the opposite outcome, even if minorities are preferred
to non-minorities and are, on average, at least as qualified.

We propose a model of search, evaluation, and selection of candidates
for a position. Consider, for example, the procedure for hiring a
new university faculty member. The evaluation of candidates is sequential
and involves a sequence of signals about the candidates' qualification,
including recommendation letters, interviews, seminars, and the evaluation
of previous research and teaching output. The hiring process is governed
by university recruitment standards, which may (but need not) include
``hard'' affirmative-action policies. We consider two types of SAA
interventions: (1) promoting the expansion of the candidate pool in
addition to evaluating those already in it, and (2) changing search
practices in order to boost the number of minority candidates in the
pool, at the expense of non-minorities. We allow for the possibility
that the university may have different standards for accepting minority
candidates and may place higher value on hiring a minority candidate.
The evaluation procedure is carried out by a recruiting committee,
which decides on the sequence of candidates\textquoteright{} evaluations
as well as when (if at all) to search for additional candidates during
the process.

We first consider the effects of SAA policies when hiring decisions
are conducted under a myopic rule (i.e., when the sequence of candidates
to evaluate and the decision to search for additional ones are based
on myopic values attached to these actions). We show that SAA policies
may reduce the probability of selecting a minority candidate, even
if (i) hiring a minority candidate is at least as valuable as hiring
a non-minority one, conditional on the candidate being qualified,
(ii) minority candidates are ex-ante at least as likely to be qualified
as non-minorities, and (iii) the probability that search delivers
a minority candidate is at least as high as the probability it brings
a non-minority one.

We then show the same conclusions apply under the optimal forward-looking
rule. The mechanism underlying our results is the following. Search
for additional candidates is an alternative to the evaluation of existing
ones. Since SAA policies alter the desirability of expanding the candidate
pool, they affect the relative desirability of evaluating candidates
already in it. Since the evaluation of minority candidates is noisier,
this may shift the evaluation of existing candidates in favor of non-minorities.
Importantly, such policies may backfire both when they increase the
relative attractiveness of search, and when they decrease it.

Consider first policies increasing the relative attractiveness of
search (e.g., through the reduction of its costs). By making search
more attractive, such policies may induce the committee to substitute
the further evaluation of candidates whose early evaluation yielded
negative results (e.g., unconvincing interviews, low test scores,
etc.) with search for new candidates. Because the evaluation of minority
candidates is noisier, the chances that negative early evaluations
are due to noise rather than lack of qualification are greater for
minority candidates, who therefore suffer from the interruption of
the evaluation process more than non-minorities. Our results show
that this mechanism may be strong enough to undermine the ultimate
chances that a minority candidate is chosen.

Next, consider policies reducing the attractiveness of search relative
to a more thorough evaluation of candidates already in the pool. An
example of such policies is the introduction of explicit requirements
to shift search from fields/schools/areas populated primarily by non-minorities
towards those populated by minorities. Because the candidates brought
in by search are expected to be more difficult to evaluate, such policies
ultimately reduce the attractiveness of search relative to a lengthier
evaluation of candidates already in the pool. We show that the reduction
in the attractiveness of search in turn may lead to a relatively more
thorough evaluation of non-minority candidates compared to minority
ones. Consequently, when the above effects are strong enough, despite
raising the percentage of minority candidates in the pool, such policies
may reduce the overall probability of hiring a minority candidate.

Our results are driven by the assumption that the evaluation of minority
candidates is noisier than that of non-minority ones. This assumption
is a central feature in the literature on statistical discrimination\footnote{In contrast to taste-based theories of discrimination (\hyperlink{Becker}{Becker, 1957}),
statistical discrimination theories explain group inequality without
assuming prejudice or preference bias.} building on \hyperlink{Phelps}{Phelps' (1972)} seminal contribution
(e.g., \hyperlink{Aigner}{Aigner and Cain, 1977}, \hyperlink{Borjas}{Borjas and Goldberg, 1979},
\hyperlink{Lundberg}{Lundberg and Straz,  1983}, \hyperlink{Cornell}{Cornell and Welch, 1996},
and more recently \hyperlink{Chambers}{Chambers and Echenique, 2018,}
and \hyperlink{Strulovici}{Bardhi et al., 2020}).\footnote{An important exception is \hyperlink{Arrow}{Arrow's (1973)} theory
of statistical discrimination, which differs from Phelps' and relies
on coordination failures rather than differences in the evaluation
of particular groups. See \hyperlink{Fang}{Fang and Moro (2011)}
for an overview of the statistical discrimination literature.} The literature has offered various explanations for the relative
difficulties in the evaluation of minority candidates . For instance,
recruiting committees may have more limited experience evaluating
minorities. \hyperlink{Aigner}{Aigner and Cain (1977)} suggest that
majority-group candidates face ``a more homogeneous set of environmental
determinants of quality,'' resulting in a lower variance in qualification,
and hence less noisy evaluation. It may be easier to assess a candidate's
history if it followed a well-known path, and it may be easier to
interpret a candidate's references when they come from familiar letter
writers. Furthermore, certain tests used in recruitment were initially
designed with a specific group in mind; for example, it has long been
argued that the SAT is more informative about the abilities of White
students than African-American students (\hyperlink{Linn}{Linn, 1973},
\hyperlink{Fleming}{Fleming and Garcia, 1998}).

The difficulty in evaluating minorities may also originate in differences
in ``background'' between minority candidates and recruiters.\footnote{``Background'' here may include race, sex, language, sexual preference,
ethnic background, religious beliefs, and nationality.} Such differences may impede the assessment of intangible qualities
such as character, charisma, responsiveness, initiative, or focus,
which are informal, but nevertheless potentially relevant for the
position (see, e.g., \hyperlink{Arrow1}{Arrow, 1972}, and \hyperlink{Cornell}{Cornell and Welch, 1996}).
Relatedly, minority candidates may use ``languages'' (broadly defined)
less familiar to the recruiting committee. \hyperlink{Lang}{Lang (1986)}
proposes a ``language theory'' of discrimination, whereby noisier
evaluation of minorities is the result of differences in ``language''
impeding communication. Recent research also suggests that women tend
to use more tentative language (qualifiers/disclaimers/hedges/intensifiers)
in job interviews than men, which may be misinterpreted as lack of
assertiveness (\hyperlink{Leaper}{Leaper and Robnett, 2011}). Similarly,
output differences due to maternity/family duties may be acknowledged
but difficult to quantify. Further, recruiters often (implicitly)
use recent hires with a similar background as a benchmark for evaluating
new candidates. If mostly white men were hired in the recent past,
such benchmarking may provide recruiters with less effective tools
for evaluating minority candidates through similarity comparisons.

Our results suggest that, without taking steps to improve the evaluation
of minority candidates, attempts to shift the composition of the candidate
pool in favor of minorities may simply amount to ``checking a box,''
or even prove detrimental. Such steps may include creating a sufficiently
diverse recruiting committee and ensuring evaluation is based on a
predetermined set of objective criteria with predetermined weights.
In our model, if differences in the evaluation of candidates were
entirely eliminated -- as in the case of ``blind'' auditions --
SAA would be guaranteed to increase the probability of hiring a minority
candidate. If differences in candidates' evaluation cannot be sufficiently
reduced, ``hard'' affirmative action may be necessary to foster
minority recruitment.

The literature on statistical discrimination has also studied the
unintended effects of affirmative action, such as its negative effects
on incentives to invest in human capital, productivity stereotyping,
wage differentials, and occupational social status (see, e.g., \hyperlink{Fryer1}{Fryer and Loury, 2005}).\footnote{The literature on the economics of affirmative action includes \hyperlink{Welch}{Welch (1976)},
\hyperlink{Lundberg}{Lundberg and Straz (1983)}, \hyperlink{Chung}{Chung (2000)},
\hyperlink{Moro}{Moro and Norman (2003)}, and \hyperlink{Fryer2}{Fryer and Loury (2013)},
and is too broad to be succinctly discussed here. For evidence on
the effectiveness of affirmative action, see \hyperlink{Holzer}{Holzer and Neumark (2000)}.} In an influential article, \hyperlink{Coate}{Coate and Loury (1993)}
show, for example, that a ``patronizing equilibrium'' may arise
in which minorities' incentives to invest in skills may be reduced
when affirmative action is in place. This paper complements this literature
by illustrating the possible negative effects of SAA directly on the
dimension it is intended for -- the enhancement of minority recruitment.

\section{Model}

The recruitment problem described below applies to a variety of environments.
For concreteness, we focus on a university committee recruiting faculty,
or students, for an available position. The recruitment of a candidate
is subject to university approval standards. There are two categories
of candidates, $A$ and $B$ (race, gender, field of study). Candidates'
qualifications $\theta\in\{L,H\}$ are unknown and independent ex-ante.
Let $p_{0}^{j}=\Pr(\theta^{j}=H)$ denote the prior that a category-$j$
candidate is qualified. The value of hiring a qualified category-$j$
candidate is $v^{j}>0$, whereas the value of hiring any non-qualified
candidate is zero. That is, the school wishes to hire only qualified
candidates, but candidates of one category may be preferred. 

Candidates can be recruited only if they are in the school's candidate
pool. For simplicity, we assume that at the outset there are only
two candidates in the pool, one from each category. At each period
$t=0,1,...,$ the committee either chooses a candidate to evaluate
among those in the pool, or decides to expand the pool by searching
for additional candidates. The evaluation of a candidate generates
new information, formally captured by a signal about the candidate's
qualification. Again, for simplicity, we assume signals are binary.
The evaluation of a qualified candidate from category $j$ yields
a signal realization $s=1$ with probability $q_{H}^{j}\equiv\text{Pr}(s=1|\theta^{j}=H)$,
whereas the evaluation of a non-qualified category-$j$ candidate
yields a signal realization $s=0$ with probability $q_{L}^{j}\equiv\text{Pr}(s=0|\theta^{j}=L)$,
with $q_{H}^{j}\geq1-q_{L}^{j}$. Conditional on a candidate's type,
the signals are iid draws from the above Bernoulli distribution.

Given a history $\sigma=(s_{1},s_{2},...)$ of signal realizations,
the posterior probability that a category-$j$ candidate is qualified
will be denoted by $p^{j}(\sigma)$. The null history is denoted by
$\sigma=\emptyset$.

Each time the committee searches for new candidates, it identifies
a category-$j$ candidate with probability $\mu^{j}$, and no candidate
with probability $1-\mu^{A}-\mu^{B}$.\footnote{The results extend to more general search technologies, in particular,
the possibility that each search yields a random number of candidates
from each category. We assume a single candidate is discovered at
each search to ease the exposition.} For simplicity, we assume that beyond the opportunity cost of not
being able to evaluate one of the candidates in the pool while searching,
there is no direct cost for search. Likewise, the only cost of evaluating
a candidate already in the pool is the cost of postponing search and
the evaluation of other candidates.

The entire recruiting process ends when the committee finds a candidate
whose (category-adjusted) expected quality is large enough. Formally,
assume that for any $j$ there exists a threshold $\bar{P}^{j}\in(0,1]$
such that each candidate from category $j$ is given the slot at history
$\sigma$ if and only if $p^{j}(\sigma)\geq\bar{P}^{j}$. The threshold
$\bar{P}^{j}$ can be thought of as reflecting standards imposed by
the university, which may, but need not, coincide with the preferences
of the recruiting committee. To avoid trivialities, assume $p_{0}^{j}<\bar{P}^{j}$,
which means each candidate must be evaluated at least once to be recruited.

A \textit{recruitment rule} specifies in each period either a candidate
to evaluate among those in the pool, or search for a new candidate.
A recruitment rule is \textit{optimal }if it maximizes the expected
discounted payoff $\mathbb{E}\left(\delta^{T}\tilde{v}_{T}\right)$
over all feasible recruitment rules, where $\delta\in(0,1)$ is the
discount factor, $T$ the (stochastic) time at which a selection is
made, and $\tilde{v}_{T}$ the value of the selected candidate.\footnote{If the selected candidate is from category $j$ and the history of
signals is $\sigma$, the expected value of the candidate is $p^{j}(\sigma)v^{j}$.}

As anticipated in the Introduction, the key assumption behind our
results is that the evaluation of minority candidates is noisier than
that of non-minority ones. Formally, $A$ is the minority category
if and only if the evaluation of $B$-candidates is Blackwell more
informative than that of $A$-candidates. Under the technology described
above, this amounts to $q_{k}^{B}\geq q_{k}^{A}$, $k=H,L$, with
at least one strict inequality.

\section{Myopic Recruitment Rule\label{sec:Myopic-Selection-Strategy}}

We start by assuming the recruiting process is conducted under a simple,
myopic rule. Given a history of signal realizations $\sigma$, denote
by $\lambda^{j}(\sigma)$ the probability that a candidate from category
$j$ with history $\sigma$ is recruited after a single additional
evaluation (that is, $\lambda^{j}(\sigma)=\text{Pr}(s:p^{j}(\sigma,s)\geq\bar{P}^{j}|j,\sigma$)).
The myopic value the committee attaches to an additional evaluation
of a category-$j$ candidate with history $\sigma$ is equal to $u^{j}(\sigma)=\lambda^{j}(\sigma)v^{j}$.
Likewise, the myopic value it attaches to an additional search is
equal to the expected value $u^{S}=\delta\left(\mu^{A}u^{A}(\emptyset)+\mu^{B}u^{B}(\emptyset)\right)$
of bringing a ``blank-slate'' candidate to the pool. Under a myopic
rule, in each period, the committee selects the alternative (evaluation
of a candidate from the pool, or search) with the highest myopic value.
To avoid trivialities, assume $u^{A}(\emptyset),u^{B}(\emptyset)>u^{S}$,
so that upon being added to the pool each candidate has a greater
myopic value than search.

Given $v\equiv(v^{A},v^{B})$, $p_{0}\equiv(p_{0}^{A},p_{0}^{B})$,
$\mu\equiv(\mu^{A},\mu^{B})$, $q\equiv(q_{L}^{A},q_{H}^{A},q_{L}^{B},q_{H}^{B})$,
and $\bar{P}\equiv(\bar{P}^{A},\bar{P}^{B})$, denote by $\gamma^{j}(v,p_{0},\mu,q,\bar{P})$
the ex-ante probability a category-$j$ candidate is selected under
a myopic rule.

Our first result pertains to SAA policies aimed at expanding the size
of the candidate pool (e.g., instructions to the committee to include
areas typically ignored by the school in its searches). Formally,
such policies can be captured by an improvement in search technology,
i.e., an increase in $\mu^{A}$ and $\mu^{B}$ (possibly asymmetric
across categories). To ease the exposition, and without any important
implication for the results, we assume that, prior to the introduction
of the policy, $\mu=(0,0)$.
\begin{prop}
\label{prop:myopic-adverse-effect-of-search}Suppose $A$ is the minority
category and that recruitment is conducted under a myopic rule. There
exist $(v,p_{0},\mu,q,\overline{P})$ such that $A$-candidates are
(a) more valuable to the school ($v^{A}>v^{B}$), (b) more likely
to be qualified ($p_{0}^{A}>p_{0}^{B}$), (c) more likely to be identified
through search $(\mu^{A}>\mu^{B}>0)$, (d) have a lower acceptance
threshold ($\bar{P}^{A}<\bar{P}^{B}$), and yet policies promoting
the expansion of the candidate pool may reduce the ex-ante probability
that $A$-candidates are selected: $\gamma^{A}(v,p_{0},\mu,q,\bar{P})<\gamma^{A}(v,p_{0},\vec{0},q,\bar{P})$.
\end{prop}
Under the assumptions in the proposition, an improvement in the search
technology naturally increases the expected number of minority candidates
considered for the position, as well as their share relative to non-minority
ones. However, the increase in $\mu^{A}$ and $\mu^{B}$ -- by making
search more attractive -- might come at the expense of lengthier
evaluations of candidates whose early evaluations yielded negative
results. Because minority candidates are more difficult to evaluate
than non-minority ones, they are the ones who are more likely to suffer
from the truncation in the evaluation process. The latter effect,
when strong enough, may imply a reduction in the ex-ante probability
of selecting a minority candidate.

Note that the probability of selecting $A$-candidates is reduced
not only relative to the probability of selecting $B$-candidates,
but overall; that is, promoting expansion of the candidate pool may
reduce the ex-ante probability of recruiting $A$-candidates, despite
increasing the overall probability of filling the slot.

Next, consider SAA policies aimed at increasing the probability of
finding $A$-candidates at the expense of $B$-candidates (formally
captured by an increase in $\mu^{A}$ and an equal reduction in $\mu^{B}$).
\begin{prop}
\label{prop:myopic-affirmative_action_may_harm}Suppose $A$ is the
minority category and that recruitment is conducted under a myopic
rule. There exist $(v,p_{0},\mu,q,\overline{P})$, with $v^{A}>v^{B}$,
$p_{0}^{A}>p_{0}^{B}$ and $\overline{P}^{A}<\overline{P}^{B}$ such
that SAA policies aimed at increasing the likelihood of finding $A$-candidates
at the expense of $B$-candidates reduce the ex-ante probability of
selecting an $A$-candidate: $\gamma^{A}\left(v,p_{0},(\mu^{A},\mu^{B}),q,\overline{P}\right)>\gamma^{A}\left(v,p_{0},(\mu^{A}+\zeta,\mu^{B}-\zeta),q,\overline{P}\right)$,
for $\zeta>0$.
\end{prop}
Under such SAA policies, $A$-candidates are more likely to be included
in the candidate pool at the expense of $B$-candidates, increasing
the probability of recruiting $A$-candidates. On the other hand,
because $A$-candidates are more difficult to evaluate than $B$-candidates,
such policies, contrary to those examined above, reduce the overall
attractiveness of search relative to a lengthier evaluation of existing
candidates. Because $B$-candidates are the easiest to evaluate, the
committee may substitute search primarily with the evaluation of the
$B$-candidates. The proof in the Appendix shows that the latter effect
may be strong enough to trigger a reduction in the ex-ante probability
that minority candidates are selected.

\section{Optimal (forward-looking) Rule}

We now show that the effects identified above are not a mere consequence
of the committee following a myopic rule -- they may emerge also
under an optimal rule.

\subsection{Preliminaries}

Consider first an environment in which the candidate pool is exogenous
and constant over time (this amounts to $\mu^{A}=\mu^{B}=0$). As
we show in the Online Appendix, the optimal rule is an index rule
(\hyperlink{Gittins}{Gittins and Jones, 1974}). Each candidate is
assigned an index $V^{j}$ that depends only on the candidate's category,
$j$, and the candidate's history of signal realizations, $\sigma$.
The indices take the following form (see the proof of Lemma \ref{lem:optimal-selection-wout-search}
in the Online Appendix)
\begin{align}
V^{j} & (\sigma)=\sup_{\tau^{j}>0}\frac{\mathbb{E}\left[\delta^{\phi^{j}}\left(1-\delta^{\tau^{j}-\phi^{j}}\right)\boldsymbol{1}_{\{\phi^{j}<\tau^{j}\}}\tilde{v}^{j}|j,\sigma\right]}{1-\mathbb{E}\left[\delta^{\tau^{j}}|j,\sigma\right]}\label{eq:index-arm}
\end{align}
where $\tau^{j}$ is a (stochastic) stopping-time, $\phi^{j}$ is
the (stochastic) time at which the candidate's (posterior) probability
of success $p^{j}(\sigma)$ exceeds the acceptance threshold $\bar{P}^{j}$
for the first time, and $\tilde{v}^{j}\in\{0,v^{j}\}$ denotes the
candidate's value to the organization. The index (\ref{eq:index-arm})
is the maximal average expected discounted payoff (averaging over
discounted time) from the sequential evaluation of the candidate,
ignoring the existence of the other candidates.
\begin{lem}
\label{lem:optimal-selection-wout-search}Assume the candidate pool
is fixed. In each period, the optimal recruitment rule evaluates the
candidate with the highest index, as defined in (\ref{eq:index-arm}).
\end{lem}
Next, consider an environment where the candidate pool can be expanded
through search. Let
\begin{align}
V^{S} & =\sup_{\tau^{A},\tau^{B}>0}\frac{\delta\sum_{j\in\{A,B\}}\mu^{j}\mathbb{E}\left[\delta^{\phi^{j}}\left(1-\delta^{\tau^{j}-\phi^{j}}\right)\boldsymbol{1}_{\{\phi^{j}<\tau^{j}\}}\tilde{v}^{j}|j,\emptyset\right]}{1-\sum_{j\in\{A,B\}}\mu^{j}\mathbb{E}\left[\delta^{\tau^{j}}|j,\emptyset\right]},\label{eq:search_index}
\end{align}
be the index of search, where $\tau^{j}$, $\phi^{j}$ and $\tilde{v}^{j}$
are as defined above. The index maximizes the average expected discounted
payoff from the sequential evaluation of the first candidate that
arrives as the result of search.
\begin{lem}
\label{lem:optimal-selection-w-search} Suppose the candidate pool
can be expanded through search. The optimal recruitment rule consists
in evaluating in each period one of the candidates in the pool with
the highest index, as defined in (\ref{eq:index-arm}), provided this
index is greater than $V^{S},$ and searching for new candidates otherwise.
\end{lem}
The fact that the optimal rule takes an index form follows from arguments
similar to those in the literature on branching bandits (e.g., \hyperlink{Weiss}{Weiss, 1988},
\hyperlink{Weber}{Weber, 1992}). In \hyperlink{FershtmanPavan}{Fershtman and Pavan (2019)},
we show that, under appropriate conditions, the problem of searching
for ``arms'' is a special case of the branching problem. In that
paper we also provide a novel proof for the optimality of an index
rule, and derive a recursive characterization of the index for search
which we use here to arrive at (\ref{eq:search_index}). In the Online
Appendix of this paper, we show that the optimality of an index rule
carries over to the recruitment problem under consideration here,
despite the irreversibility of the recruitment decisions.\footnote{In general, index policies are not guaranteed to be optimal in the
presence of irreversible decisions.}

Under the optimal rule, the value the committee assigns to the additional
evaluation of any candidate in the pool takes into account the effects
of the results of the additional evaluation on the desirability of
all further evaluations. Likewise, the value the committee assigns
to search accounts for the fact that the new candidates may be evaluated
multiple times. Note that $V^{S}$ is directly linked to the indices
of the candidates expected to be identified by search. Specifically,
the optimal stopping-time $\tau$ in (\ref{eq:search_index}) is the
first time at which the index of any candidate that arrives as a result
of search drops below $V^{S}$.

\subsection{Soft affirmative action}

The following result shows that the negative effects of improvements
in search technology on the recruitment of minorities may occur also
under optimal (forward-looking) policies. Let $\Gamma^{j}(v,p_{0},\mu,q,\overline{P})$
denote the ex-ante probability of selecting a category-$j$ candidate
under the optimal rule.
\begin{prop}
\label{prop:optimal-adverse-effect-of-search} Suppose $A$ is the
minority category and that recruitment is conducted under an optimal
rule. There exist $(v,p_{0},\mu,q,\overline{P})$ with $v^{A}>v^{B}$,
$p_{0}^{A}>p_{0}^{B}$, $\overline{P}^{A}<\overline{P}^{B}$ and $\mu^{A}>\mu^{B}>0$
such that SAA policies promoting the expansion of the candidate pool
may reduce the ex-ante probability that $A$-candidates are selected:
$\Gamma^{A}(v,p_{0},\mu,q,\overline{P})<\Gamma^{A}(v,p_{0},\vec{0},q,\overline{P})$.
\end{prop}
The intuition for the result is similar to the one for the myopic
rule. Qualified minority candidates whose initial evaluation yields
negative results due to noise may not have the opportunity to prove
themselves when search is an attractive alternative to further evaluation.
Although improvements in the search technology may increase the presence
of minority candidates in the pool, they may reduce the ex-ante probability
that the position is given to a minority candidate. Importantly, this
may happen despite the fact that improvements in the search technology
may bring more minority candidates to the pool than non-minority ones.

Similarly, policies that increase $\mu^{A}$ at the expense of $\mu^{B}$
may also backfire by reducing the ex-ante probability that $A$-candidates
are selected.
\begin{prop}
\label{prop:optimal-affirmative-action} Suppose $A$ is the minority
category and that recruitment is conducted under an optimal rule.
There exist $(v,p_{0},\mu,q,\overline{P})$, with $v^{A}>v^{B}$,
$p_{0}^{A}>p_{0}^{B}$, and $\overline{P}^{A}<\overline{P}^{B}$,
such that SAA policies increasing the likelihood that search brings
$A$-candidates at the expense of $B$-candidates reduce the ex-ante
probability of selecting an $A$-candidate: $\Gamma^{A}\left(v,p_{0},(\mu^{A},\mu^{B}),q,\overline{U}\right)>\Gamma^{A}\left(v,p_{0},(\mu^{A}+\zeta,\mu^{B}-\zeta),q,\overline{U}\right)$,
for $\zeta>0$.
\end{prop}
The mechanism behind the result in Proposition \ref{prop:optimal-affirmative-action}
is similar to the one under the myopic rule. Such policies reduce
the overall attractiveness of search relative to a more careful evaluation
of the candidates already in the pool. Because, among those candidates
already in the pool, $A$-candidates are the most difficult to evaluate,
the committee may substitute search with the evaluation of the $B$-candidates
at the expense of the $A$-candidates. When this effect is strong
enough, such policies may have the unintended effect of reducing the
ex-ante probability of selecting a minority candidate.

\section{Discussion\label{sec:Implications-for-implementing}}

The Rooney rule was adopted in 2003. Yet in 2020, there are only three
African-American head coaches, the same number as in 2003, prompting
criticism viewing the rule as merely a means of ``checking a box''.\footnote{See the \textit{Washington Post}, Jan 5, 2020: ``The dearth of black
coaches in the NFL is a problem that somehow still hasn\textquoteright t
been fixed''.} 

Why are certain SAA policies unsuccessful? This paper suggests a possible
explanation based on differences in the effectiveness of evaluating
minority and non-minority candidates, which does not presume any bias
in decision making. SAA policies do increase the percentage of minority
candidates considered for a position, but also alter the desirability
of searching for more candidates relative to evaluating those already
in the pool. The purpose of this article is to show that this effect
may be strong enough to lead to a reduction in the overall probability
that a minority candidate is selected. Unaccompanied by steps to reduce
difficulties in the evaluation of minority candidates, SAA is thus
not guaranteed to deliver its desired effects; it may indeed amount
to ``checking a box,'' or in some cases even prove counterproductive.

\newpage{}

\section*{A$\ \ $Proofs}

\textbf{Proof of Proposition \ref{prop:myopic-adverse-effect-of-search}.
$\ $ }Suppose $q_{H}^{A}\in(0,1)$, and $q_{L}^{A}=q_{H}^{B}=q_{L}^{B}=1$.
That is, the evaluation of $B$-candidates is perfectly revealing,
whereas the evaluation of $A$-candidates takes the typical ``no-news-is-bad-news''
form. Under this technology, $\lambda^{B}(\emptyset)=p_{0}^{B}$,
$\lambda^{A}(\emptyset)=q_{H}^{A}p_{0}^{A}$, and $\lambda^{A}(0)=q_{H}^{A}p^{A}(0)$,
with
\begin{align*}
p^{A}(0) & =\frac{(1-q_{H}^{A})p_{0}^{A}}{(1-q_{H}^{A})p_{0}^{A}+1-p_{0}^{A}}.
\end{align*}

\emph{Status-quo technology} ($\mu^{A}=\mu^{B}=0$). The $A$-candidate
is evaluated first if 
\begin{align}
u^{A}(\emptyset)=p_{0}^{A}q_{H}^{A}v^{A} & >p_{0}^{B}v^{B}=u^{B}(\emptyset).\label{eq:myopic_A_first}
\end{align}
Assume (\ref{eq:myopic_A_first}) holds. Then $\gamma^{A}(v,p_{0},\vec{0},q,\overline{P})=p_{0}^{A}\left(1-p_{0}^{B}+p_{0}^{B}q_{H}^{A}\right)$.

\emph{Improved search technology} ($\mu^{A},\mu^{B}>0$ with $\mu^{B}=1-\mu^{A}$).
Condition (\ref{eq:myopic_A_first}), along with the assumption that
$u^{B}(\emptyset),u^{A}(\emptyset)>u^{S}$, implies that
\begin{align}
\underbrace{p_{0}^{A}q_{H}^{A}v^{A}}_{u^{A}(\emptyset)}>\underbrace{p_{0}^{B}v^{B}}_{u^{B}(\emptyset)} & >\underbrace{\delta\left(\mu^{A}p_{0}^{A}q_{H}^{A}v^{A}+\mu^{B}p_{0}^{B}v^{B}\right)}_{u^{S}}.\label{eq:myopic-B-preferred-to-search-initially}
\end{align}
Now suppose that
\begin{align}
\underbrace{\frac{(1-q_{H}^{A})p_{0}^{A}q_{H}^{A}v^{A}}{(1-q_{H}^{A})p_{0}^{A}+1-p_{0}^{A}}}_{u^{A}(0)} & <\delta\left(\mu^{A}p_{0}^{A}q_{H}^{A}v^{A}+\mu^{B}p_{0}^{B}v^{B}\right).\label{eq:myopic_search_better_than_A_after_single_fail}
\end{align}
Condition (\ref{eq:myopic_search_better_than_A_after_single_fail})
implies a single negative evaluation of an $A$-candidate suffices
to trigger search. Denote by $\gamma_{S}^{A}$ the probability of
selecting an $A$-candidate after search is carried-out. Under a myopic
rule, when search is carried-out, any existing candidate is never
evaluated again. Therefore, 
\begin{align*}
 & \mathbb{\gamma}_{S}^{A}=\mu^{A}\left(\lambda^{A}(\emptyset)+(1-\lambda^{A}(\emptyset))\gamma_{S}^{A}\right)+\mu^{B}(1-p_{0}^{B})\gamma_{S}^{A}.
\end{align*}
 Rearranging and using $\mu^{A}+\mu^{B}=1$, we have that
\begin{align*}
 & \mathbb{\mathbb{\gamma}}_{S}^{A}=\frac{\mu^{A}\lambda^{A}(\emptyset)}{\mu^{A}\lambda^{A}(\emptyset)+\mu^{B}p_{0}^{B}}.
\end{align*}
 Hence,
\begin{align*}
\gamma^{A}(v,p_{0},\mu,q) & =\lambda^{A}(\emptyset)+(1-\lambda^{A}(\emptyset))(1-p_{0}^{B})\mathbb{\gamma}_{S}^{A}=p_{0}^{A}q_{H}^{A}\left(1+\frac{\mu^{A}(1-\lambda^{A}(\emptyset))(1-p_{0}^{B})}{\mu^{A}\lambda^{A}(\emptyset)+\mu^{B}p_{0}^{B}}\right).
\end{align*}
Given (\ref{eq:myopic_A_first})-(\ref{eq:myopic_search_better_than_A_after_single_fail}),
$\gamma^{A}(v,p_{0},\mu,q,\overline{P})<\gamma^{A}(v,p_{0},\vec{0},q,\overline{P})$
if 
\begin{align*}
 & p_{0}^{A}\left(1-p_{0}^{B}+p_{0}^{B}q_{H}^{A}\right)>p_{0}^{A}q_{H}^{A}\left(1+\frac{\mu^{A}(1-\lambda^{A}(\emptyset))(1-p_{0}^{B})}{\mu^{A}\lambda^{A}(\emptyset)+\mu^{B}p_{0}^{B}}\right),
\end{align*}
or, equivalently, 
\begin{align}
\mu^{B}p_{0}^{B}(1-q_{H}^{A}) & >\mu^{A}(1-p_{0}^{A})q_{H}^{A}.\label{eq:eq:Myopic_prob_A_reduced_w_search}
\end{align}

The result in the proposition follows by observing that there exists
a non-empty open set of parameter values satisfying both the restrictions
in the proposition and Conditions (\ref{eq:myopic_A_first})-(\ref{eq:eq:Myopic_prob_A_reduced_w_search})
(the following is an example: $\delta=0.9$, $p_{0}^{A}=0.8$, $p_{0}^{B}=0.7$,
$v^{A}=1.5$, $v^{B}=1$, $q_{H}^{A}=0.6$, $\mu^{A}=2/3$). $\hfill\blacksquare$

\bigskip{}

\noindent \textbf{Proof of Proposition \ref{prop:myopic-affirmative_action_may_harm}.
$\ $ }Let $q_{L}^{A}=q_{L}^{B}=1$ and $1>q_{H}^{B}>q_{H}^{A}>0$.
Then $\lambda^{j}(\emptyset)=p_{0}^{j}q_{H}^{j}$, $p^{j}(0)=(1-q_{H}^{j})p_{0}^{j}/\left(1-q_{H}^{j}p_{0}^{j}\right)$,
and $\lambda^{j}(0)=q_{H}^{j}p^{j}(0)$. Also assume $\mu^{B}=1-\mu^{A}.$

\emph{Status-quo technology} ($\zeta=0$) Suppose that{\small{}
\begin{align}
\underbrace{\lambda^{B}(\emptyset)v^{B}}_{u^{B}(\emptyset)} & >\underbrace{\lambda^{A}(\emptyset)v^{A}}_{u^{A}(\emptyset)}>\underbrace{\delta\left(\mu^{A}\lambda^{A}(\emptyset)v^{A}+\mu^{B}\lambda^{B}(\emptyset)v^{B}\right)}_{u^{S}}>\underbrace{\frac{(1-q_{H}^{A})\lambda^{A}(\emptyset)v^{A}}{1-q_{H}^{A}p_{0}^{A}}}_{u^{A}(0)},\underbrace{\frac{(1-q_{H}^{B})\lambda^{B}(\emptyset)v^{B}}{1-q_{H}^{B}p_{0}^{B}}}_{u^{B}(0)}.\label{eq:zeta=00003D0}
\end{align}
}Condition (\ref{eq:zeta=00003D0}) implies that evaluating any blank-slate
candidate is preferred to search, whereas search is preferred to evaluating
any candidate whose first evaluation yielded a negative result.

\emph{Affirmative action in search} ($\zeta>0$). Assume
\begin{align}
 & \frac{(1-q_{H}^{B})\lambda^{B}(\emptyset)v^{B}}{1-q_{H}^{B}p_{0}^{B}}>\delta\left((\mu^{A}+\zeta)\lambda^{A}(\emptyset)v^{A}+(\mu^{B}-\zeta)\lambda^{B}(\emptyset)v^{B}\right)>\frac{(1-q_{H}^{A})\lambda^{A}(\emptyset)v^{A}}{1-q_{H}^{A}p_{0}^{A}}.\label{eq:zeta>0}
\end{align}
Note that (\ref{eq:zeta>0}) implies that, after a single negative
evaluation, search is preferred to a second evaluation if the candidate
is a $B$-candidate, whereas the opposite holds for $A$-candidates.

\emph{Comparison}. Denote by $\gamma_{S}^{A}(z)$, $z\in\{0,\zeta\}$,
the probability of selecting an $A$-candidate after search is launched,
with $z=0$ in case of the default technology, and $z=\zeta$ under
SAA.

Under the myopic rule, once search is carried out, any candidate already
in the pool is never evaluated again. Therefore, 
\begin{align*}
 & \mathbb{\gamma}_{S}^{A}(0)=\mu^{A}\left(\lambda^{A}(\emptyset)+(1-\lambda^{A}(\emptyset))\gamma_{S}^{A}(0)\right)+\mu^{B}\left(1-\lambda^{B}(\emptyset)\right)\gamma_{S}^{A}(0),
\end{align*}
which implies
\begin{align*}
 & \mathbb{\gamma}_{S}^{A}(0)=\frac{\mu^{A}\lambda^{A}(\emptyset)}{\mu^{A}\lambda^{A}(\emptyset)+\mu^{B}\lambda^{B}(\emptyset)}.
\end{align*}

It follows that 
\begin{align*}
\mathbb{\mathbb{\gamma}}^{A}\left(v,p_{0},(\mu^{A},\mu^{B}),q,\overline{P}\right) & =(1-\lambda^{B}(\emptyset))\left(\lambda^{A}(\emptyset)+(1-\lambda^{A}(\emptyset))\mathbb{\gamma}_{S}^{A}(0)\right)\\
 & =(1-\lambda^{B}(\emptyset))\lambda^{A}(\emptyset)\left(\frac{\mu^{A}+\mu^{B}\lambda^{B}(\emptyset)}{\mu^{A}\lambda^{A}(\emptyset)+\mu^{B}\lambda^{B}(\emptyset)}\right).
\end{align*}
Similarly, Conditions (\ref{eq:zeta=00003D0})-(\ref{eq:zeta>0})
imply that
\begin{align*}
\mathbb{\gamma}_{S}^{A}(\zeta) & <(\mu^{A}+\zeta)\left(\lambda^{A}(\emptyset)+(1-\lambda^{A}(\emptyset))\gamma_{S}^{A}(\zeta)\right)+(\mu^{B}-\zeta)(1-\lambda^{B}(\emptyset))(1-\lambda^{B}(0))\gamma_{S}^{A}(\zeta),
\end{align*}
where the inequality follows from the fact that a $B$-candidate may
potentially be evaluated more than twice before search is launched.
Rewriting the above inequality, we have that
\begin{align*}
\mathbb{\gamma}_{S}^{A}(\zeta) & <\frac{(\mu^{A}+\zeta)\lambda^{A}(\emptyset)}{(\mu^{A}+\zeta)\lambda^{A}(\emptyset)+(\mu^{B}-\zeta)\left(\lambda^{B}(\emptyset)+\lambda^{B}(0)\left(1-\lambda^{B}(\emptyset)\right)\right)}.
\end{align*}
Therefore, 
\begin{align*}
 & \mathbb{\mathbb{\gamma}}^{A}\left(v,p_{0},(\mu^{A}+\zeta,\mu^{B}-\zeta),q,\overline{P}\right)\\
 & \ <(1-\lambda^{B}(\emptyset))\left(\lambda^{A}(\emptyset)+(1-\lambda^{A}(\emptyset))(1-\lambda^{B}(0))\mathbb{\gamma}_{S}^{A}(\zeta)\right)\\
 & \ =(1-\lambda^{B}(\emptyset))\lambda^{A}(\emptyset)\left(1+\frac{(1-\lambda^{B}(0))(1-\lambda^{A}(\emptyset))(\mu^{A}+\zeta)}{(\mu^{A}+\zeta)\lambda^{A}(\emptyset)+(\mu^{B}-\zeta)\left(\lambda^{B}(\emptyset)+\lambda^{B}(0)\left(1-\lambda^{B}(\emptyset)\right)\right)}\right).
\end{align*}
Hence, $\mathbb{\mathbb{\gamma}}^{A}\left(v,p_{0},(\mu^{A}+\zeta,\mu^{B}-\zeta),q,\overline{P}\right)<\mathbb{\mathbb{\gamma}}^{A}\left(v,p_{0},(\mu^{A},\mu^{B}),q,\overline{P}\right)$
if
\begin{align}
\frac{\mu^{A}}{\mu^{A}\lambda^{A}(\emptyset)+\mu^{B}\lambda^{B}(\emptyset)} & >\frac{(1-\lambda^{B}(0))(\mu^{A}+\zeta)}{(\mu^{A}+\zeta)\lambda^{A}(\emptyset)+(\mu^{B}-\zeta)\left(\lambda^{B}(\emptyset)+\lambda^{B}(0)\left(1-\lambda^{B}(\emptyset)\right)\right)}.\label{eq:myopic_prob_A_smaller_with_zeta_increase}
\end{align}

The result in the proposition follows by observing that there exists
a non-empty and open set of parameter values satisfying both the restrictions
in the proposition and Conditions (\ref{eq:zeta=00003D0})-(\ref{eq:myopic_prob_A_smaller_with_zeta_increase})
(for example: $\delta=0.9$, $p_{0}^{A}=0.8$, $p_{0}^{B}=0.64$,
$v^{A}=1.5$, $v^{B}=1$, $q_{H}^{A}=0.2$, $q_{H}^{B}=0.75$, $\mu^{A}=0.9$,
and $\zeta=0.04$). $\hfill\blacksquare$

\bigskip{}

\noindent \textbf{Proof of Proposition \ref{prop:optimal-adverse-effect-of-search}.
}Consider the same evaluation technology as in Proposition \ref{prop:myopic-adverse-effect-of-search}:
$q_{H}^{A}\in(0,1)$, and $q_{L}^{A}=q_{H}^{B}=q_{L}^{B}=1$. The
index of a blank-slate $j$-candidate is then equal to
\begin{align}
V^{j}(\emptyset) & =\frac{\lambda^{j}(\emptyset)v^{j}}{1-\delta(1-\lambda^{j}(\emptyset))},\label{eq:index_for_A_and_B}
\end{align}
as the optimal $\tau^{j}$ in the definition of the index specifies
stopping immediately ($\tau^{j}=1$) following a negative signal $s=0$,
and never stopping ($\tau^{j}=\infty$) following a positive signal
$s=1$. Recall that, under the assumed evaluation technology, $\lambda^{B}(\emptyset)=p_{0}^{B}$
and $\lambda^{A}(\emptyset)=p_{0}^{A}q_{H}^{A}$. Furthermore, $V^{B}(0)=0$
and $V^{j}(1)=v^{j}$. It is also easy to see that, analogously to
(\ref{eq:index_for_A_and_B}), 
\begin{align*}
 & V^{A}(0)=\frac{\lambda^{A}(0)v^{j}}{1-\delta(1-\lambda^{A}(0))}.
\end{align*}

\emph{Status-quo technology} ($\mu^{A}=\mu^{B}=0$). In this case,
$V^{S}=0$. The $B$-candidate is then evaluated first if
\begin{align}
V^{B}(\emptyset)=\frac{p_{0}^{B}v^{B}}{1-\delta(1-p_{0}^{B})} & >\frac{p_{0}^{A}q_{H}^{A}v^{A}}{1-\delta(1-p_{0}^{A}q_{H}^{A})}=V^{A}(\emptyset).\label{eq:optimal_B_evaluated_before_A}
\end{align}
Assuming (\ref{eq:optimal_B_evaluated_before_A}) holds, $\Gamma^{A}(v,p_{0},\vec{0},q,\overline{P})=(1-p_{0}^{B})p_{0}^{A}.$

\emph{Improved search technology} ($\mu^{A},\mu^{B}>0$, with $\mu^{B}=1-\mu^{A}$
). From Lemma \ref{lem:optimal-selection-w-search}, the optimal stopping-time
$\tau$ in (\ref{eq:search_index}) is the first time at which the
index of the candidate discovered through search drops below $V^{S}$.
Now suppose
\begin{align}
V^{A}(\emptyset),V^{B}(\emptyset) & >\frac{\delta\left(\mu^{A}p_{0}^{A}q_{H}^{A}v^{A}+\mu^{B}p_{0}^{B}v^{B}\right)}{1-\delta^{2}+\delta^{2}\left(\mu^{A}p_{0}^{A}q_{H}^{A}+\mu^{B}p_{0}^{B}\right)}>V^{A}(0).\label{eq:indices_initially_greater_than_search}
\end{align}

We argue that the index of search is then equal to 
\begin{align}
V^{S} & =\frac{\delta\left(\mu^{A}p_{0}^{A}q_{H}^{A}v^{A}+\mu^{B}p_{0}^{B}v^{B}\right)}{1-\delta^{2}+\delta^{2}\left(\mu^{A}p_{0}^{A}q_{H}^{A}+\mu^{B}p_{0}^{B}\right)}.\label{eq:eq:index_search_specific_evaluation_structure}
\end{align}
To see this, recall that the optimal stopping-times $\tau^{j}$ in
(\ref{eq:search_index}) are the first times at which the index of
the arriving candidate drops weakly below $V^{S}$. Now suppose this
time coincides with the first time at which an evaluation yields a
negative signal realization $s=0$, and else is equal to $\tau^{j}=+\infty$.
Using these observations, (\ref{eq:eq:index_search_specific_evaluation_structure})
follows from the definition of the search index. Condition (\ref{eq:indices_initially_greater_than_search}),
along with the fact that $V^{B}(0)=0$, guarantee the optimal stopping-times
$\tau^{j}$ in (\ref{eq:search_index}) are indeed the ones assumed
above.

Given Lemma \ref{lem:optimal-selection-w-search} and the stationarity
of the search index, if search is preferred to the evaluation of a
candidate at some history, it continues to be preferred to the evaluation
of that candidate in all future periods. 

Now let $\Gamma_{S}^{A}$ denote the probability of selecting an $A$-candidate
once search is launched. Given the above conditions,
\begin{align*}
\Gamma_{S}^{A}=\mu^{A}\left(\lambda^{A}(\emptyset)+(1-\lambda^{A}(\emptyset))\Gamma_{S}^{A}\right)+\mu^{B}\left(1-p_{0}^{B}\right)\Gamma_{S}^{A},
\end{align*}
 which yields
\begin{align*}
 & \Gamma_{S}^{A}=\frac{\mu^{A}\lambda^{A}(\emptyset)}{\mu^{A}\lambda^{A}(\emptyset)+\mu^{B}p_{0}^{B}}.
\end{align*}

Therefore, the ex-ante probability of selecting an $A$-candidate
is equal to
\begin{align*}
\Gamma^{A}(v,p_{0},\mu,q,\overline{P}) & =(1-p_{0}^{B})\left(\lambda^{A}(\emptyset)+(1-\lambda^{A}(\emptyset))\Gamma_{S}^{A}\right)\\
 & =(1-p_{0}^{B})\lambda^{A}(\emptyset)\left(\frac{\mu^{A}+\mu^{B}p_{0}^{B}}{\mu^{A}\lambda^{A}(\emptyset)+\mu^{B}p_{0}^{B}}\right).
\end{align*}
Under Conditions (\ref{eq:optimal_B_evaluated_before_A})-(\ref{eq:indices_initially_greater_than_search}),
$\Gamma^{A}(v,p_{0},\mu,q,\overline{P})<\Gamma^{A}(v,p_{0},\vec{0},q,\overline{P})$
if
\begin{align*}
(1-p_{0}^{B})p_{0}^{A} & >(1-p_{0}^{B})\lambda^{A}(\emptyset)\left(\frac{\mu^{A}+\mu^{B}p_{0}^{B}}{\mu^{A}\lambda^{A}(\emptyset)+\mu^{B}p_{0}^{B}}\right).
\end{align*}
The latter condition is equivalent to Condition (\ref{eq:eq:Myopic_prob_A_reduced_w_search}).
The result in the proposition now follows by observing that there
exists a non-empty open set of parameter values satisfying the restrictions
in the proposition, Conditions (\ref{eq:optimal_B_evaluated_before_A})-(\ref{eq:indices_initially_greater_than_search}),
and Condition (\ref{eq:eq:Myopic_prob_A_reduced_w_search}) (for example:
$\delta=0.9$, $p_{0}^{A}=0.75$, $p_{0}^{B}=0.7$, $v^{A}=1.2$,
$v^{B}=1$, $q_{H}^{A}=0.19$, and $\mu^{A}=0.52$). $\hfill\blacksquare$\bigskip{}

\noindent \textbf{Proof of Proposition \ref{prop:optimal-affirmative-action}.}
Consider the same evaluation technology as in the proof of Proposition
\ref{prop:myopic-affirmative_action_may_harm} (i.e., $q_{L}^{A}=q_{L}^{B}=1$,
and $1>q_{H}^{B}>q_{H}^{A}>0$) and assume $\mu^{B}=1-\mu^{A}$. The
same arguments as in the proof of Proposition \ref{prop:optimal-adverse-effect-of-search}
imply that $V^{j}(\emptyset)=\lambda^{j}(\emptyset)v^{j}/\left(1-\delta(1-\lambda^{j}(\emptyset))\right)$
and $V^{j}(0)=\lambda^{j}(0)v^{j}/\left(1-\delta(1-\lambda^{j}(0))\right)$,
with $\lambda^{j}(\emptyset)=p_{0}^{j}q_{H}^{j}$ and 
\begin{align*}
 & \lambda^{j}(0)=p^{j}(0)q_{H}^{j}=\frac{(1-q_{H}^{j})p_{0}^{j}q_{H}^{j}}{1-q_{H}^{j}p_{0}^{j}}.
\end{align*}

Assume
\begin{align}
\frac{p_{0}^{B}q_{H}^{B}v^{B}}{1-\delta(1-p_{0}^{B}q_{H}^{B})} & >\frac{p_{0}^{A}q_{H}^{A}v^{A}}{1-\delta(1-p_{0}^{A}q_{H}^{A})}\label{eq:prop4_optimal_B_evaluated_before_A}
\end{align}
and observe that the latter condition implies $V^{B}(\emptyset)>V^{A}(\emptyset)$,
so that the $B$-candidate is evaluated first before search is launched.

\emph{Status-quo technology} ($\zeta=0$). Assume
\begin{equation}
V^{A}(\emptyset),V^{B}(\emptyset)>\frac{\delta\left(\mu^{A}p_{0}^{A}q_{H}^{A}v^{A}+\mu^{B}p_{0}^{B}q_{H}^{B}v^{B}\right)}{1-\delta^{2}+\delta^{2}\left(\mu^{A}p_{0}^{A}q_{H}^{A}+\mu^{B}p_{0}^{B}q_{H}^{B}\right)}>V^{A}(0),V^{B}(0),\label{eq:prop4_standard}
\end{equation}
and note that the above condition implies that the optimal stopping-time
in the formula for $V^{S}$ is the first time at which an evaluation
yields a realization $s=0$ which in turn implies that
\begin{equation}
V^{S}=\frac{\delta\left(\mu^{A}p_{0}^{A}q_{H}^{A}v^{A}+\mu^{B}p_{0}^{B}q_{H}^{B}v^{B}\right)}{1-\delta^{2}+\delta^{2}\left(\mu^{A}p_{0}^{A}q_{H}^{A}+\mu^{B}p_{0}^{B}q_{H}^{B}\right)}.\label{eq:new-formula}
\end{equation}

\emph{Affirmative action in search} ($\zeta>0$). Assume {\small{}
\begin{equation}
V^{A}(\emptyset),V^{B}(0)>\frac{\delta(\mu^{A}+\zeta)\lambda^{A}(\emptyset)v^{A}+\delta(\mu^{B}-\zeta)v^{B}\left(\lambda^{B}(\emptyset)+\delta(1-\lambda^{B}(\emptyset))\lambda^{B}(0)\right)}{1-\delta^{2}\left((\mu^{A}+\zeta)\left(1-\lambda^{A}(\emptyset)\right)+(\mu^{B}-\zeta)(1-\lambda^{B}(\emptyset))(1-\lambda^{B}(0))\delta\right)}>V^{A}(0),V^{B}(0,0),\label{eq:prop4_two_failures}
\end{equation}
}where 
\begin{align*}
 & V^{B}(0,0)=\frac{\lambda^{B}(0,0)v^{B}}{1-\delta(1-\lambda^{B}(0,0))},
\end{align*}
 with $\lambda^{B}(0,0)=p^{B}(0,0)q_{H}^{B}$ and 
\begin{align*}
 & p^{B}(0,0)=\frac{(1-q_{H}^{B})p^{B}(0)}{1-q_{H}^{B}p^{B}(0)}.
\end{align*}
Condition (\ref{eq:prop4_two_failures}) implies that it takes one
negative signal realization $s=0$ for the index of an $A$-candidate
to drop below 
\begin{align}
V^{S}=\frac{\delta(\mu^{A}+\zeta)\lambda^{A}(\emptyset)v^{A}+\delta(\mu^{B}-\zeta)v^{B}\left(\lambda^{B}(\emptyset)+\delta(1-\lambda^{B}(\emptyset))\lambda^{B}(0)\right)}{1-\delta^{2}\left((\mu^{A}+\zeta)\left(1-\lambda^{A}(\emptyset)\right)+(\mu^{B}-\zeta)(1-\lambda^{B}(\emptyset))(1-\lambda^{B}(0))\delta\right)}\label{eq:eq:prop4_index_search_specific_evaluation_structure}
\end{align}
and two negative signal realizations $s=0$ for the index of a $B$-candidate
to drop below the value of $V^{S}$ in (\ref{eq:eq:prop4_index_search_specific_evaluation_structure}),
thus implying that the search index is indeed the value in (\ref{eq:eq:prop4_index_search_specific_evaluation_structure}).

Condition (\ref{eq:myopic_prob_A_smaller_with_zeta_increase}) in
the proof of Proposition \ref{prop:myopic-affirmative_action_may_harm}
then implies
\[
\Gamma^{A}\left(v,p_{0},(\mu^{A},\mu^{B}),q,\overline{P}\right)>\Gamma^{A}\left(v,p_{0},(\mu^{A}+\zeta,\mu^{B}-\zeta),q,\overline{P}\right).
\]
The result in the proposition then follows by observing that there
exists a non-empty open set of parameter values satisfying both the
restrictions in the proposition and Conditions (\ref{eq:prop4_optimal_B_evaluated_before_A}),
(\ref{eq:prop4_standard}), (\ref{eq:prop4_two_failures}), and (\ref{eq:myopic_prob_A_smaller_with_zeta_increase})
(e.g., $\delta=0.9$, $p_{0}^{A}=0.69$, $p_{0}^{B}=0.68$, $v^{A}=1.01$,
$v^{B}=1$, $q_{H}^{A}=0.4$, $q_{H}^{B}=0.8$, $\mu^{A}=0.15$, and
$\zeta=0.01$). $\hfill\blacksquare$

\newpage{}
\begin{center}
\textbf{\LARGE{}Online Appendix}{\LARGE\par}
\par\end{center}

\appendix\setcounter{section}{1}

\section{Omitted proofs}

\noindent \textbf{Proof of Lemma \ref{lem:optimal-selection-wout-search}.
$\ $} In the absence of search, the recruitment problem can be mapped
into a multi-armed bandit problem, with candidates corresponding to
different ``arms''. This problem, however, is not a standard one,
because of the irreversibility of the recruiting decision. Notwithstanding
the fact that many multi-arm bandit problems with irreversible choice
fail to admit an index solution, the arguments below imply that an
index rule is optimal in our model.

To see this, consider the following fictitious environment. Let the
flow payoff from the pull of each arm be equal to zero at each pull
for which $p^{j}(\sigma)<\bar{P}^{j}$ (such pulls correspond to the
evaluation of candidates which, given the university's acceptance
rules, are not acceptable yet). Once $p^{j}(\sigma)\geq\bar{P}^{j}$,
the pull of the arm generates a constant flow payoff equal to $p^{j}(\sigma^{\phi^{j}})v^{j}(1-\delta)$
in each subsequent period, and the ``state'' of the arm does not
change any more (here $\sigma^{\phi^{j}}$ is the history of signal
realizations at the first time at which the candidate's probability
of being qualified exceeds the acceptance threshold $\bar{P}^{j}$).
Further assume that at any point in time the committee can pull any
arm -- even if for some candidate $p^{j}(\sigma)\geq\bar{P}^{j}$
as the result of previous pulls. In other words, in contrast to the
true model, the pull of an arm corresponding to a candidate for whom
$p^{j}(\sigma)\geq\bar{P}^{j}$ does not preclude the possibility
of pulling other arms in subsequent periods. This environment satisfies
all the conditions guaranteeing the optimality of an index rule in
the classic multi-armed bandit problem. Therefore, the optimal rule
in such fictitious environment selects in each period the arm with
the highest Gittins index. The Gittins index of each candidate is
equal to
\begin{align*}
V^{j}(\sigma) & =\text{sup}_{\tau^{j}>0}\frac{\mathbb{E}\left[\sum_{s=\phi^{j}}^{\tau^{j}-1}\delta^{s}v^{j}p^{j}(\sigma^{\phi^{j}})(1-\delta)|j,\sigma\right]}{1-\mathbb{E}\left[\delta^{\tau^{j}}|j,\sigma\right]}\\
 & =\text{sup}_{\tau^{j}>0}\frac{\mathbb{E}\left[\delta^{\phi^{j}}\left(1-\delta^{\tau^{j}-\phi^{j}}\right)\boldsymbol{1}_{\{\phi^{j}<\tau^{j}\}}\tilde{v}^{j}|j,\sigma\right]}{1-\mathbb{E}\left[\delta^{\tau^{j}}|j,\sigma\right]}
\end{align*}

Next note that, once an arm is pulled for which $p^{j}(\sigma)\geq\bar{P}^{j}$,
its index $V^{j}(\sigma)=p^{j}(\sigma^{\phi^{j}})v^{j}(1-\delta)$
remains the same at all subsequent periods, and the indices of the
other arms remain the same. Hence, under the optimal rule in the fictitious
environment, once a candidate is found qualified (i.e., $p^{j}(\sigma)\geq\bar{P}^{j}$),
that candidate is selected in each subsequent period. Because the
fictitious environment is a relaxation of the primitive one, it follows
that the optimal rule in the primitive environment coincides with
the one in the fictitious environment. $\hfill\blacksquare$\bigskip{}

\noindent \textbf{Proof of Lemma \ref{lem:optimal-selection-w-search}.}
$\ $ In the presence of search, the recruitment problem can be thought
of as a generalization of the classic multi-armed bandit problem in
which the decision maker can search for new arms, in addition to pulling
one of the existing ones. As shown in \hyperlink{FershtmanPavan}{Fershtman and Pavan (2019)}
(see also the literature on branching bandits), in the absence of
irreversible decisions, the optimal rule for such problems continues
be an index rule, but with a special index for search. The latter
index is the maximal expected average discounted payoff from searching
or pulling any of the new arms that arrive as the result of search,
where the maximization is over both a stopping time and a rule that
chooses among search and the pull of the new arms that arrive as the
result of search. Arguments similar to those in the proof of Lemma
\ref{lem:optimal-selection-wout-search} imply that, notwithstanding
the irreversibility of the recruiting decision, the optimal rule for
the recruiting problem under consideration is an index rule. The results
in \hyperlink{FershtmanPavan}{Fershtman and Pavan (2019)} also imply
that the search index is equal to
\begin{align}
V^{S} & =\sup_{\tau,\pi}\frac{\delta\mathbb{E}^{\pi,\tau}\left[\sum_{s=1}^{\tau-1}\delta^{s}r_{s}^{\pi}\right]}{1-\mathbb{E}^{\pi,\tau}\left[\delta^{\tau}\right]},\label{eq:index_search_general}
\end{align}
where $\tau$ is a stopping time, $\pi$ is a rule that chooses between
evaluation of one of the new candidates brought in by search and further
search, and $r_{s}^{\pi}$ is the flow payoff under the rule $\pi$,
with the latter equal to zero when $\pi$ selects search or one of
the candidates for which $p^{j}(\sigma)<\bar{P}^{j}$, and is equal
to $p^{j}(\sigma^{\phi^{j}})v^{j}(1-\delta)$ when $\pi$ selects
a candidate for whom, either at the present period or at a previous
period, $p^{j}(\sigma)\geq\bar{P}^{j}$ (recall that $\sigma^{\phi^{j}}$
is the history of signal realizations at the first time at which the
candidate's probability of being qualified exceeds the acceptance
threshold).

Importantly, note that the index of search is independent of any information
pertaining to the candidates already in the candidate pool.

It can further be shown that the optimal rule $\pi$ in (\ref{eq:index_search_general})
is in fact the same index rule that characterizes the solution to
the entire problem, and that the optimal stopping time $\tau$ in
(\ref{eq:index_search_general}) is the first time at which the search
index and the index of all newly arrived arms fall below the index
of search when the latter was launched. In the context of the specific
problem under examination here, because the search technology is stationary,
$\tau$ coincides with the first time at which the index of the first
candidate identified by search falls below the index of search. Given
these features, (\ref{eq:index_search_general}) can be rewritten
as
\begin{align*}
V^{S} & =\sup_{\tau^{A},\tau^{B}>0}\frac{\delta\sum_{j\in\{A,B\}}\mu^{j}\mathbb{E}\left[\sum_{s=\phi^{j}}^{\tau^{j}-1}\delta^{s}p^{j}(\sigma^{\phi^{j}})v^{j}(1-\delta)|j,\emptyset\right]}{1-\sum_{j\in\{A,B\}}\mu^{j}\mathbb{E}\left[\delta^{\tau^{j}}|j,\emptyset\right]}\\
 & =\sup_{\tau^{A},\tau^{B}>0}\frac{\delta\sum_{j\in\{A,B\}}\mu^{j}\mathbb{E}\left[\delta^{\phi^{j}}\left(1-\delta^{\tau^{j}-\phi^{j}}\right)\boldsymbol{1}_{\{\phi^{j}<\tau^{j}\}}\tilde{v}^{j}|j,\emptyset\right]}{1-\sum_{j\in\{A,B\}}\mu^{j}\mathbb{E}\left[\delta^{\tau^{j}}|j,\emptyset\right]}.
\end{align*}
 $\hfill\blacksquare$\bigskip{}


\begin{thebibliography}{10}
\bibitem{key-1}\textbf{\hypertarget{Aigner}{}Aigner, D. and G. Cain}
(1977). ``Statistical Theories of Discrimination in Labor Markets,''\textit{
Industrial and Labor Relations Review, 30}(2), 175-187.

\bibitem{key-2}\textbf{\hypertarget{Arrow1}{}Arrow, K. }(1972).
``Models of Job Discrimination,'' In \textit{Racial Discrimination
in Economic Life}.

\bibitem{key-1}\textbf{\hypertarget{Arrow}{}Arrow, K.} (1973). \textit{The
Theory of Discrimination}, \textit{in Orley Ashenfelter and Albert
Rees, eds., Discrimination in Labor Markets}, Princeton, NJ: Princeton
University Press, pp. 3-33.

\bibitem{key-3}\textbf{\hypertarget{Strulovici}{}Bardhi, A., Guo,
Y. and B. Strulovici} (2020). ``Spiraling or Self-Correcting Discrimination:
A Multi-Armed Bandit Approach,'' Working paper.

\bibitem{key-2}\textbf{\hypertarget{Becker}{}Becker, G. S.} (1957).
\textit{The Theory of Discrimination}, University of Chicago Press.

\bibitem{key-2}\textbf{\hypertarget{Borjas}{}Borjas, G. J. and M.
S. Goldberg,} (1979). ``Biased Screening and Discrimination in the
Labor Market,\textquotedbl{} \textit{American Economic Review, 68}(5),
918-22.

\bibitem{key-3}\textbf{\hypertarget{Chambers}{}Chambers, C. P. and
F. Echenique} (2018). ``A Characterization of ``Phelpsian'' Statistical
Discrimination,'' Working paper, arXiv preprint arXiv:1808.01351.

\bibitem{key-7-1}\textbf{\hypertarget{Chung}{}Chung, K.-S.} (2000).
``Role Models and Arguments for Armative Action,'' \textit{The American
Economic Review}, \textit{90}, No. 3, 640-648.

\bibitem{key-4-1}\textbf{\hypertarget{Coate}{}Coate, S. and G. C.
Loury,} (1993). ``Will Affirmative-Action Policies Eliminate Negative
Stereotypes?,'' \textit{The American Economic Review,} \textit{83}(5),
1220-1240.

\bibitem{key-4}\textbf{\hypertarget{Cornell}{}Cornell, B. and I.
Welch }(1996). ``Culture, Information, and Screening Discrimination,''
\textit{Journal of Political Economy}, \textit{104}(3), 542-571.

\bibitem{key-5}\textbf{\hypertarget{DuBois}{}DuBois, C.} (2015).
``The Impact of \textquotedblleft Soft\textquotedblright{} Affirmative
Action Policies on Minority Hiring in Executive Leadership: The Case
of the NFL's Rooney Rule,'' \textit{American Law and Economics Review,
18}(1), 208-233.

\bibitem{key-6}\textbf{\hypertarget{Fang}{}Fang, H. and A. Moro}
(2011). ``Theories of Statistical Discrimination and Affirmative
Action: A Survey,'' In\textit{ Handbook of Social Economics} (Vol.
1, pp. 133-200). North-Holland.

\bibitem{key-6}\textbf{\hypertarget{FershtmanPavan}{}Fershtman,
D. and A. Pavan} (2019). ``Sequential Learning with Endogenous Consideration
Sets,'' Working paper, Tel Aviv University and Northwestern University.

\bibitem{key-5}\textbf{\hypertarget{Fleming}{}Fleming, J. and N.
Garcia} (1998). ``Are Standardized Tests Fair to African Americans?
Predictive Validity of the SAT in Black and White Institutions,''
\textit{The Journal of Higher Education, 69}(5), 471-495.

\bibitem{key-1}\textbf{\hypertarget{Fryer1}{}Fryer, R. and G. C.
Loury }(2005). \textquotedblleft Affirmative Action and Its Mythology,\textquotedblright{}
\textit{Journal of Economic Perspectives 19} (3): 147--62.

\bibitem{key-6}\textbf{\hypertarget{Fryer2}{}Fryer, R. and G. C.
Loury} (2013). ``Valuing Diversity,'' \textit{Journal of Political
Economy 121}(4), 747-774.

\bibitem{key-7}\textbf{\hypertarget{Gittins}{}Gittins, J. and D.
Jones} (1974). ``A Dynamic Allocation Index for the Sequential Design
of Experiments,'' In J. Gani (Ed.), \textit{Progress in Statistics},
pp. 241-266. Amsterdam, NL: North- Holland.

\bibitem{key-2}\textbf{\hypertarget{Heilman}{}Heilman, M. E.} (1980).
\textquotedblleft The Impact of Situational Factors on Personnel Decisions
Concerning Women: Varying the Sex Composition of the Applicant Pool,\textquotedblright{}
\textit{Organizational Behavior and Human Performance 26}, 286--295.

\bibitem{key-4}\textbf{\hypertarget{Holzer}{}Holzer, H. and D. Neumark}
(2000). ``Assessing Affirmative Action,'' \textit{Journal of Economic
Literature, 38}(3), 483-568.

\bibitem{key-8}\textbf{\hypertarget{Lang}{}Lang, K.} (1986). ``A
Language Theory of Discrimination,'' \textit{The Quarterly Journal
of Economics}, \textit{101}(2), 363-382.

\bibitem{key-1}\textbf{\hypertarget{Leaper}{}Leaper, C. and R. D.
Robnett }(2011). ``Women Are More Likely than Men to Use Tentative
Language, Aren't They? A Meta-Analysis Testing for Gender Differences
and Moderators,'' \textit{Psychology of Women Quarterly, 35}(1) 129-142.

\bibitem{key-1}\textbf{\hypertarget{Linn}{}Linn R. L.} (1973). ``Fair
Test Use in Selection,'' \textit{Review of Educational Research},
Vol. 43, No. 2, pp. 139-61.

\bibitem{key-9}\textbf{\hypertarget{Lundberg}{}Lundberg, S. and
R. Startz} (1983). ``Private Discrimination and Social Intervention
in Competitive Labor Market,'' \textit{The American Economic Review,
73}(3), 340-347.

\bibitem{key-3}\textbf{\hypertarget{Moro}{}Moro, A. and P. Norman}
(2003). ``Affirmative Action in a Competitive Economy,'' \textit{Journal
of Public Economics, 87}(3-4), 567-594.

\bibitem{key-10}\textbf{\hypertarget{Phelps}{}Phelps, E.} (1972).
``The Statistical Theory of Racism and Sexism,'' \textit{The American
Economic Review, 62}(4), 659-661.

\bibitem{key-11}\textbf{\hypertarget{Schuck}{}Schuck, P. H.} (2002).
``Affirmative Action: Past, Present, and Future,'' \textit{Yale
Law \& Policy Review, 20}, 1.

\bibitem{key-12}\textbf{\hypertarget{Weber}{}Weber, R.} (1992).
``On the Gittins Index for Multiarmed Bandits,'' \textit{The Annals
of Applied Probability, 2}(4), 1024-1033.

\bibitem{key-13}\textbf{\hypertarget{Weiss}{}Weiss, G.} (1988).
``Branching Bandit Processes,'' \textit{Probability in the Engineering
and Informational Sciences, 2}(3), 269-278.

\bibitem{key-2}\textbf{\hypertarget{Welch}{}Welch, F.} (1976). ``Employment
Quotas for Minorities,'' \textit{Journal of Political Economy, 84}(4,
Part 2), S105-S141.
\end{thebibliography}
\end{document}